\documentclass[twocolumn,prl,superscriptaddress]{revtex4}
\usepackage[colorlinks,linkcolor=blue,anchorcolor=blue,citecolor=blue,filecolor=blue,menucolor=blue,runcolor=blue,urlcolor=blue,frenchlinks=blue]{hyperref}
\usepackage{amsmath}
\usepackage{graphicx}
\usepackage{amssymb}
\usepackage{dcolumn}
\usepackage{mathrsfs}
\usepackage{bm}
\begin{document}

\title{{Measurement of non-Abelian gauge fields using multi-loop amplification}}

\author{Qing-Xian Lv}
\affiliation{Key Laboratory of Atomic and Subatomic Structure and Quantum Control (Ministry of Education), School of Physics, South China Normal University, Guangzhou 510006, China}

\affiliation{Guangdong Provincial Key Laboratory of Quantum Engineering and Quantum Materials, Guangdong-Hong Kong Joint Laboratory of Quantum Matter, South China Normal University, Guangzhou 510006, China}

\author{Hong-Zhi Liu}
\affiliation{Key Laboratory of Atomic and Subatomic Structure and Quantum Control (Ministry of Education), School of Physics, South China Normal University, Guangzhou 510006, China}

\author{Yan-Xiong Du}
\email{yanxiongdu@m.scnu.edu.cn}
\affiliation{Key Laboratory of Atomic and Subatomic Structure and Quantum Control (Ministry of Education), School of Physics, South China Normal University, Guangzhou 510006, China}

\author{Lin-Qing Chen}
\affiliation{Key Laboratory of Atomic and Subatomic Structure and Quantum Control (Ministry of Education), School of Physics, South China Normal University, Guangzhou 510006, China}

\author{Meng Wang}
\affiliation{Key Laboratory of Atomic and Subatomic Structure and Quantum Control (Ministry of Education), School of Physics, South China Normal University, Guangzhou 510006, China}

\author{Jia-Hao Liang}
\affiliation{Key Laboratory of Atomic and Subatomic Structure and Quantum Control (Ministry of Education), School of Physics, South China Normal University, Guangzhou 510006, China}

\author{Zhao-Xin Fu}
\affiliation{Key Laboratory of Atomic and Subatomic Structure and Quantum Control (Ministry of Education), School of Physics, South China Normal University, Guangzhou 510006, China}

\author{Zi-Yuan Chen}
\affiliation{Key Laboratory of Atomic and Subatomic Structure and Quantum Control (Ministry of Education), School of Physics, South China Normal University, Guangzhou 510006, China}

\author{Hui Yan}
\email{yanhui@scnu.edu.cn}
\affiliation{Key Laboratory of Atomic and Subatomic Structure and Quantum Control (Ministry of Education), School of Physics, South China Normal University, Guangzhou 510006, China}

\affiliation{Guangdong Provincial Key Laboratory of Quantum Engineering and Quantum Materials, Guangdong-Hong Kong Joint Laboratory of Quantum Matter, South China Normal University, Guangzhou 510006, China}

\affiliation{Guangdong Provincial Engineering Technology Research Center for Quantum Precision Measurement, Frontier Research Institute for Physics, South China Normal University, Guangzhou 510006, China}

\author{Shi-Liang Zhu}
\email{slzhu@scnu.edu.cn}
\affiliation{Key Laboratory of Atomic and Subatomic Structure and Quantum Control (Ministry of Education), School of Physics, South China Normal University, Guangzhou 510006, China}

\affiliation{Guangdong Provincial Key Laboratory of Quantum Engineering and Quantum Materials, Guangdong-Hong Kong Joint Laboratory of Quantum Matter, South China Normal University, Guangzhou 510006, China}


\begin{abstract}
Non-Abelian gauge field (NAGF) plays a central role in understanding the geometrical and topological phenomena in physics. Here we experimentally induce a NAGF  in the degenerate eigen subspace of a double-$\Lambda$ four-level atomic system. The non-Abelian nature of the gauge field is detected through the measurement of the non-commutativity of two successive evolution loops.
 Then we theoretically propose and experimentally demonstrate a novel scheme to measure the NAGF through multi-loop evolution and  robust holonomic quantum gates. The demonstrated scheme offers the advantage of detecting the NAGF with amplification through multi-loop evolution. Our results pave the way for an experimentally-feasible approach to achieving high-resolution and high-precision measurements of the gauge fields.
\end{abstract}

 \maketitle

{\sl Introduction.--} The coupling between quantum systems and gauge fields gives rise to a plethora of physical phenomena, including topological insulators and topological semimetals in condensed matter systems, as well as the Aharonov-Bohm (AB) effect and geometric phases in quantum systems~\cite{Hasan2010,XQi2011,DWZhang2018,Berry1994}. Novel physics involving NAGFs, such as the $\theta$-vacuum and the non-Abelian AB effect, have also been discovered~\cite{Jackiw1976,Jackiw1980,Wu1975,Li2022}. To rigorously study these theories, it is necessary to manipulate the degenerate wave functions under the non-Abelian gauge potentials~\cite{Wilzeck1984,Duan2001,Abdumalikov,Zu2014,Toyoda2013,Leroux2018,Sj2012}. Additionally, accurately measuring the localized gauge fields, which remains a challenging task, is crucial for investigating the geometrical and topological phenomena in physics.

Linear response is one of the methods to measure the geometric quantum tensor in parameter space~\cite{Gritsev2012,Schroer,Roushan2014,XTan2021,XTan2019,Gianfrate2020,MYu2020,XTan2018}. The imaginary component of this tensor is the gauge field (Berry curvature), while the real component is the metric~\cite{SLZhu2008}. By quenching the parameters, the deflection of the evolution trajectory can be used to measure the gauge field. This approach has been successfully demonstrated in cold atomic systems, including the non-Abelian version~\cite{Kolodrubetz2016,Sugawa2018}. The linear response method offers the advantage of enabling the determination of field strength in a small quench region, thereby facilitating high-resolution measurements. However, breaking adiabaticity is necessary to achieve observable effects, which weakens the robustness against deviations in the control parameters. Cyclic evolution is another  method for measuring gauge fields and it is resistant to both random noise and systematic errors, owing to the geometric nature of geometric phases. By traversing a closed path in parameter space, a quantum system acquires a geometric phase that is determined by the integral of the Berry curvature~\cite{Berry1994,Tycko1987,Leek2007,Leroux2018,Yang2019,XDZhang2008}.  However, it has low resolution in parameter space since the enclosed area should be sufficient large to accumulate observable effect. The non-commutative Wilzeck-Zee  phases are measured in cold atomic systems \cite{Sugawa2021} and in classical system constructed by non-commutative optical elements \cite{Yang2019}. However, the local non-Abelian field can not be retrieved from the Wilzeck-Zee phase in common cases due to the path-ordered integral. Therefore, precise measurement of NAGFs is still lack of experimental study.

\begin{figure*}[ptb]
\begin{center}
\includegraphics[width=15cm]{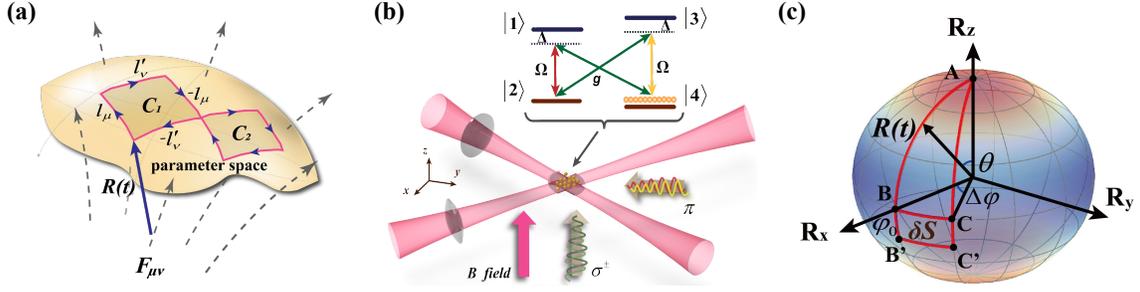}
\label{fig:set up}
\caption{\label{fig:set up} (a) Measurement of NAGFs through the use of small loops in parameter space. (b) The experimental setup and the coupling scheme. The double-$\Lambda$ four-level Hamiltonian is realized in a cold $^{87}$Rb atomic system: $|1\rangle=|F=2,m_F=-1\rangle$, $|2\rangle=|F=1,m_F=-1\rangle$, $|3\rangle=|F=2,m_F=0\rangle$ and $|4\rangle=|F=1,m_F=0\rangle$. (c) The measurement of NAGFs in square loop of B-B'-C'-C-B is carried out by analyzing the difference in evolution between two triangles A-B-C-A and A-B'-C'-A.}
\end{center}
\end{figure*}


In this Letter, we report an experiment on measurement of NAGFs with a double-$\Lambda$ four-level cold atoms.  In contrast to the conventional tripod systems~\cite{SLZhu2009,Juzeliunas2008,Ruseckas2005,DWZhang2012}, the degenerate subspace utilized in our experiments corresponds to the system's lowest eigenvalue, resulting in a longer coherence time for the exploration of NAGFs. We demonstrate that NAGFs can be measured from multi-loop evolution with tiny loop in the parameter space since the multi-loop evolution can enhance the accuracy of measurement results, enabling measurements to be taken within a smaller loop while maintaining the same level of precision. We significantly simplify the process of measuring NAGFs  in parameter space. There is a gap between the use of dressed states in defining the gauge fields and measurement of only bare states. This presents a significant challenge in experimental settings. We devise a scheme to measure the gauge field by utilizing the difference between two triangle loops that enclose the square loop. Both triangle loops originate and terminate at the pole of the Bloch sphere, and the dressed states match the bare states at the beginning and end of the evolution. Consequently, the projective measurement of dressed states can be converted to bare states. We establish a connection between holonomic quantum gates and NAGFs, and thus our method is robust against systematic errors and random noise. Therefore, our work provides a practical experimental method for attaining accurate and precise measurements of  gauge fields.

{\sl NAGFs in tiny loops.--} We begin by addressing how to measure NAGFs with cyclic evolution. Considering a Hamiltonian $H(\mathbf{R}(t))$, with a two-degenerate eigen subspace labelled by $\{|D_1\rangle, |D_2\rangle\}$, the NAGFs are given by $F_{\mu\nu}=\partial_\mu A_\nu-\partial_\nu A_\mu-[A_\mu, A_\nu]$, where the gauge potential $A^{jk}_\mu=\langle D_j|\partial_\mu|D_k\rangle\ (j, k=1, 2)$, and $\mu, \nu$ are the components of driving parameter $\mathbf{R}(t)$ \cite{Wilzeck1984}. When the system controlled by the Hamiltonian $H(\mathbf{R}(t))$ evolves adiabatically along a closed path, the evolution operator $U$ is given by a path-ordered integral
\begin{equation}
U=1-\int_0^sA_\mu d\mu+\int_0^s\int_0^{s'}A_\mu A_\nu d\mu d\nu+O(\mu^3),
\end{equation}
where $s, s' (s'\leq s)$ are upper bounds of integral.
Here, we have omitted a global dynamical phase under the adiabatic assumption~\cite{Wilzeck1984,XDZhang2008}. By considering a small loop $\mathbf {R}(t)\rightarrow\mathbf {R}(t)+\mathbf{l}_\mu\rightarrow\mathbf {R}(t)+\mathbf{l}_\mu+\mathbf{l}'_\nu\rightarrow\mathbf{l}_\mu+\mathbf{l}'_\nu\rightarrow\mathbf {R}(t)$ in the parameter space, as shown in Fig.\ref{fig:set up}(a), we obtain
\begin{equation}
U'\approx 1-F_{\mu\nu}\delta S,
\end{equation}
using the second-order approximation, where $\delta S=(\mathbf{l}_\mu \mathbf{l}'_\nu-\mathbf{l}'_\nu \mathbf{l}_\mu)/2$ is the enclosed area \cite{Polyakov}. Hence, measuring the evolution operator of tiny loops allows us to obtain the gauge fields.

\begin{figure*}[ptb]
\begin{center}
\includegraphics[width=15cm]{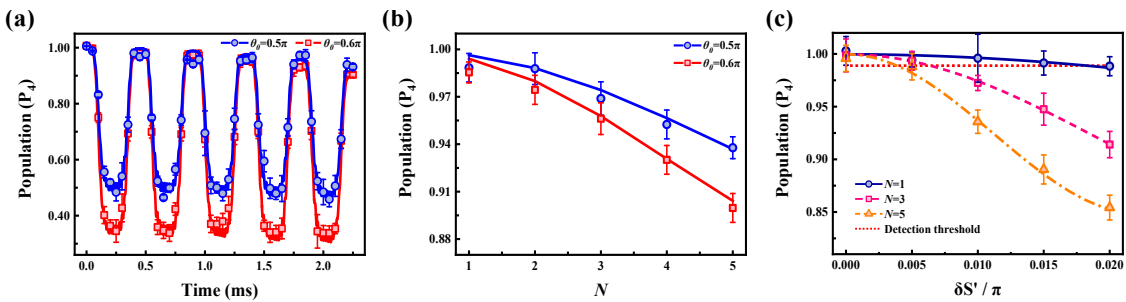}
\caption{The performance of the multi-loop approach. (a) The population $P_4$ in state $|4\rangle$ for $N=5$. The data has been average over 10 times. (b) The final population $P_4$ for $N=1$ to $N=5$. The data has been average for 20 times.  The amplification effect is demonstrated since the population decreases with $N$. (c) Minimal enclosing area that can be detected.  The threshold of minimal population deviation is set to  0.01 according to the limitation of the error bars calculation from numerous measurements. The minimal enclosing area decrease as the numbers of loops increase where the minimal enclosing area of $N=5$ loops is 4 times smaller than that of $N=1$ loop.}
\end{center}
\end{figure*}

{\sl NAGFs in a four-level system.--} NAGFs can be induced in a four-level system through a double-$\Lambda$ configuration with four microwaves, as shown in Fig.\ref{fig:set up}(b). The coupling Rabi frequencies between $|1\rangle, |2\rangle$ and $|3\rangle, |4\rangle$ are set to $\Omega$, while those between $|1\rangle, |3\rangle$ and $|2\rangle, |4\rangle$ are set to $g$. Under the bare state basis $\{|1\rangle, |2\rangle, |3\rangle, |4\rangle\}$ and rotating-wave approximation, the Hamiltonian reads
\begin{equation}
H(t)=\frac{\hbar}{2}\left ( \begin{array}{cccc}
\Delta&\Omega e^{-i\varphi}&0& -g\\
\Omega e^{i\varphi}& -\Delta & -g &0\\
0&-g& \Delta &-\Omega e^{i\varphi}\\
-g&0&-\Omega e^{-i\varphi} &  -\Delta
\end{array}\right),
\end{equation}
where $\Delta$ are the detunings defined by the difference between the microwave frequencies and the corresponding coupling levels and $\varphi$ are the relative phases~\cite{Lv2021}.
  We here have adopted the four-photon resonant condition~\cite{Lv2021}. The Hamiltonian (3) yields two two-fold degenerate eigenvalues: $\lambda_1=\lambda_2=-\Omega_0$ ($=-\sqrt{\Omega^2+g^2+\Delta^2}$) with the corresponding eigenstates ${|D_1\rangle, |D_2\rangle}$, and $\lambda_3=\lambda_4=\Omega_0$ with the corresponding eigenstates ${|B_1\rangle, |B_2\rangle}$.
  We parameterize $\Omega=\Omega_0\sin\theta\cos\phi$, $g=\Omega_0\sin\theta\sin\phi$ and $\Delta=\Omega_0\cos\theta$.
  It is important to note that the degenerate eigenstates previously proposed to generate NAGFs in a tripod level atoms  do not correspond to the lowest energy level~\cite{SLZhu2009,Juzeliunas2008,Ruseckas2005,DWZhang2012}. In contrast, we here demonstrate that NAGFs $F_{\mu \nu}$, where $\mu, \nu=\theta, \phi, \varphi$, can be obtained from the lower eigenstates ${|D_1\rangle, |D_2\rangle}$.  We will focus on the gauge field with parameters $\theta$ and $\varphi$. We note that $F_{\theta\varphi}$ is associated with the infinitesimal $d\theta d\varphi$ in Cartesian coordinates.  When we switch to spherical coordinates with infinitesimal area $\sin\theta d\theta d\varphi$, the NAGF is given by
\begin{equation}
F^S_{\theta\varphi}=F_{\theta\varphi}/\sin\theta=-i\cos\phi\mathbf{\kappa}_{\theta\varphi}\cdot\mathbf{\sigma}/2,
\end{equation}
which is obviously an SU(2) monopole. Here $\mathbf{\sigma}=(\sigma_x, \sigma_y, \sigma_z)$ with $\sigma_{x,y,z}$ being the Pauli Matrixes and  the unit vector $\mathbf{\kappa}_{\theta\varphi}=(\sin\phi\cos\varphi, \sin\phi\sin\varphi, \cos\phi)$.

The non-Abelian nature of gauge fields can be identified by the non-commutativity of two successive loops in parameter space, as shown by $C_1$ and $C_2$ in Fig.1(a). If we denote the evolution operator of $C_1$ ($C_2$) as $U_1$ ($U_2$), the evolution operators of composite paths with order $\textrm{C}_1\textrm{C}_2$ (counter-order $\textrm{C}_2\textrm{C}_1$) will be given by $U_{\mathrm{o}}=U_2U_1$ ($U_{\mathrm{co}}=U_1U_2$). In the Supplemental Material (SM)~\cite{SM}, we demonstrate that the inequality $U_2U_1\neq U_1U_2$ can be verified  through experimental data obtained by measuring non-diagonal matrix elements $|U_{\mathrm{o}}^{12(21)}|^2\neq|U_{\mathrm{co}}^{12(21)}|^2$~\cite{XDZhang2008}. These  matrix elements are determined by the population transfers after evolution and can be detected in our experiments.  As a result, we have experimentally confirmed the existence of a true NAGF in this four-level atomic system.

{\sl Detecting NAGFs in square loops.--}  As shown in Fig.1(c), we present our method for measuring the gauge field within the region $\delta S'=\Delta\theta \Delta\varphi$ enclosed by the square loop B-B'-C'-C-B.  However, it should be noted that the gauge fields are defined in the eigen subspace $\{ |D_1\rangle, |D_2\rangle \}$, which can not be directly measured. Only the population of the bare states $\{|1\rangle, |2\rangle, |3\rangle, |4\rangle\}$ can be experimentally determined.  To solve this problem, we employ  two triangle loops (as labelled by A-B-C-A and A-B'-C'-A in Fig.1(c)) to extract the NAGF in the loop B-B'-C'-C-B, with the following parameter settings: $\theta(t): 0\rightarrow\theta_0\rightarrow0$; $\varphi(t): \varphi_0\rightarrow\varphi_0+\Delta\varphi$.  In the SM~\cite{SM}, we demonstrate that the eigenstates coincide with the bare states at the starting and ending points, based on the aforementioned parameter settings. As a result, the tomography of the eigenstates can be transformed to that of the bare states. Additionally, due to the gauge potentials along the longitude line always being vanishing and not contributing to the line integral in Eq. (1) (i.e., $A_\theta=0$ \cite{SM}),  one can select appropriate paths with a sufficiently large area to accumulate a considerable observable effect. In this case, the gauge field is given as
\begin{equation}
\bar{F}_{\theta\varphi}(\theta_0, \varphi_0)=\left ( \begin{array}{cc}
\bar{F}_{\theta\varphi}^{11} & \bar{F}_{\theta\varphi}^{12}\\
\bar{F}_{\theta\varphi}^{21} & \bar{F}_{\theta\varphi}^{22}
\end{array}\right)\approx \frac{U_2-U_1}{\delta S'},
\end{equation}
where $U_{1}$ $(U_{2})$ is the evolution operator of the triangle loop A-B-C-A (A-B'-C'-A). The second-order approximation in Eq.(2) is valid when the  angle $\Delta \varphi \ll \pi$.

{\sl Amplifying the effect of NAGFs by multi-loop approach.--} The distinction between  $U_1$ and $U_2$ may not be detectable  if the gauge field within the loop B-B'-C'-C-B is not sufficient large. However, a significant advantage of the current method is that we can enhance the influence of the gauge field by iterating the progression of the loop A-B-C-A (A-B'-C'-A),  resulting in a more precise measurement.
By utilizing the multiplicity, the evolution operator for $N$ loops can be expressed as $U_\mathrm{N}=(U')^N$. Hence, if $U_\mathrm{N}$ can be measured, the evolution operator for a single loop can be derived as $U'=U_\mathrm{N}^{1/N}$, enabling the retrieval of NAGFs. Furthermore, the evolution operator of the triangle loop can be analytically obtained by  solving the Sch\"{o}rdinger's equation, which is given by
\begin{equation}
U^\mathrm{g}=\left ( \begin{array}{cc}
\cos\kappa e^{i\beta} & i\sin\kappa e^{-i\beta}\\
i\sin\kappa e^{i\beta} & \cos\kappa e^{-i\beta}
\end{array}\right),
\end{equation}
where $\kappa=\cos\phi\sin\phi\oint\cos^2(\theta/2)d\varphi=\sin(\theta_0/2)\cos\phi\sin\phi\Delta\varphi$ and $\beta=\cos^2\phi\oint\cos^2(\theta/2)d\varphi=\sin(\theta_0/2)\cos^2\phi\Delta\varphi$.  It can be found that $U'=U^g$  when $\Delta\varphi \ll \pi$~\cite{SM}. Furthermore, $U^\mathrm{g}$ is a geometric gate utilized in adiabatic Holonomic quantum computation~\cite{Unanyan1998,Zhu2002,Du2017}, which rotates the quantum state along a specific axis in the Hilbert space~\cite{SM}.
As a result, the $N$th root of the evolution operator $U_N$ can be comprehended as extracting a rotation angle from $N$ consecutive rotations. We thus establish a connection between holonomic quantum gates and NAGFs, and in the SM~\cite{SM}, we show our method is resilient to certain systematic errors and random noise.

{\sl Experimental control and the amplification effect.--}
We conducted experiments on a four-level atomic system, as depicted in Fig. 1(b). The atoms are confined within an optical dipole trap and cooled to a temperature of 10 $\mu$K via evaporation. A magnetic field is applied along the z-axis to establish the quantization axis, resulting in a frequency difference of 700 kHz between the Zeeman levels. The experimental parameters are chosen to be $\Omega_0=2\pi\times50$ kHz and $T=450 \mu$s, satisfying the adiabatic condition $\Omega_0T=45\pi\gg 2\pi$. To introduce intermediate coupling $g$, we kept $\phi=\pi/8$. The Allen-Eberly scheme \cite{Vitanov1996} was used to drive $\theta(t)$, which has the following form:
\begin{equation}
\theta(t)= \left\{\begin{split}
&\theta_0\{1+\tanh[b(t-T/4)]\}/2, t\in[0, T/2)\\
&\theta_0\{1-\tanh[b(t-3T/4)]\}/2, t\in[T/2, T]
 \end{split}\right.
\end{equation}
where $b=10/T$. $\Delta\varphi=0.1\pi$ is chosen to meet the second order approximation~\cite{SM}.

We will now discuss the amplification effect resulting from multi-loop evolution. The experiments are conducted as follows: firstly, an initial state of $|\Psi\rangle_i=|4\rangle$ is prepared. The system is then driven to evolve along the closed path A-B-C-A (A-B'-C'-A) and the final population $P_4$ remaining in $|4\rangle$ is measured and the results are plotted in Fig. 2(a).
The experimental data agree well with the theoretical calculations.
Although the control parameter $\theta(t)$  returns to the initial value at the end,  the population does not revert to its initial state. This is due to the non-Abelian geometric phase induced in the degenerate subspace.
The population decline is more rapid when $\theta=0.6\pi$ compared to $\theta=0.5\pi$ because the former has a larger enclosed area.
 In our experiments, the standard deviation approaches 0.011 after 20 iterations of averaging. To better visualize the amplification effect of multi-loop evolution, we have plotted the population of $P_4$ at the end of each loop in Fig.2(b).
 For the case of $\theta_0=0.5\pi$, the value of $P_4$ at the end of the first loop ($N=1$) is almost indistinguishable from 1. However, as the loops progress, $P_4$ decreases to $0.94(3)$ by the fifth loop, which is easily detectable in our experimental system. Similar results are  observed for the case of $\theta_0=0.6\pi$. Hence, the impact of NAGFs can be magnified through a multi-loop approach. Nonetheless, the maximum number of loops is constrained by the coherent time and flight time of the atomic system upon release. In our particular system, the Zeeman levels have a  coherence time of approximately 4 milliseconds. Consequently, we have established a maximum of 5 cycles, in accordance with the evolution period of $T=450 \mu s$.

 We subsequently address the minimum detectable area in our experiments. Given the spherically symmetric nature of the NAGF and the fact that the enclosed flux is independent of both position and shape, we specify the triangle loop (A-B-C-A in Fig. 1(c)) to discuss this issue.
 By setting $\theta_0=0.5\pi$ and scanning $\Delta\varphi$, one can tilt the enclosed area $\delta S'$. The criterion for determining the minimal area can be determined by the minimal detectable population difference of $P_4$, which is 0.011 according to the standard derivation \cite{SM}.  When preparing the system to the state $|4\rangle$, the population $P_4$  versus $\delta S'$ with $N=1, 3, 5$ is plotted in Fig. 2(c). While the minimum area required for a single loop ($N=1$) is approximately $0.2\pi$, the minimum area required for five loops ($N=5$) is only about $0.05\pi$, which is four times smaller than the former. Increasing the number of loops in the evolution process can therefore improve the measurement resolution of gauge fields.

\begin{figure}[ptb]
\begin{center}
\includegraphics[width=8cm]{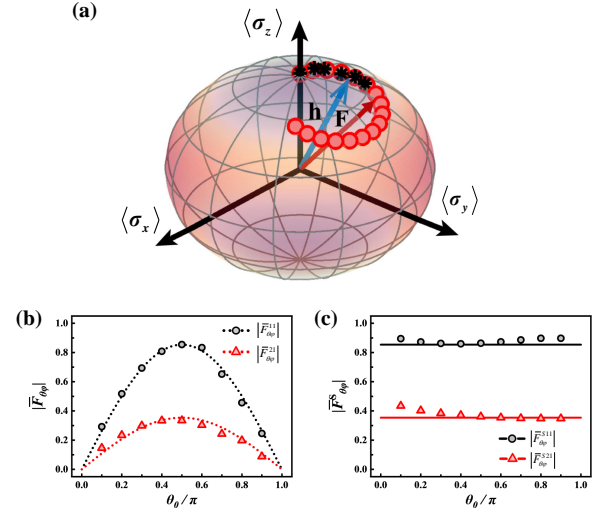}%
\caption{(a) Tomographic results of pseudospin $\mathbf{F}$ at the end of $N$ loops. Black stars: experimental results of $N=1$ to $N=5$. Red circles: theoretical simulation of $N=1$ to $N=17$ loops. Pseudospin $\mathbf{F}$ rotates along the axis $\mathbf{h}$ determined by the geometric gate $U_g$. (b) Experimental results of NAGFs. Black-dotted line with circles: $|\bar{F}_{\theta\varphi}^{11}|$. Red-dotted line with triangles: $|\bar{F}_{\theta\varphi}^{12}|$. The experimental data have been average for 20 measurements and the error bars are too small to be shown.  (c) The refined gauge field in the spherical  coordinates. Black-solid line with black circles: $|\bar{F}_{\theta\varphi}^{S11}|$. Red-solid line with red triangles: $|\bar{F}_{\theta\varphi}^{S12}|$.}
\end{center}
\end{figure}

{\sl Measurement results of NAGFs.--} We will now describe a state tomography method to extract the NAGFs. Since parameter $\theta(t)$ evolves adiabatically with $\theta(0)=\theta(T)=0$, the initial state and the final state are restricted inside the subspace spanned by $\{|2\rangle, |4\rangle\}$, which can be regarded as a pseudospin $\mathbf{F}$~\cite{Final_Population}.  We denote the final state as $|\Psi\rangle_f=a_1|2\rangle+b_1|4\rangle$ ($|\Psi'\rangle_f=a_2|2\rangle+b_2|4\rangle$) when the initial state is $|\Psi\rangle_i=|2\rangle$ ($|\Psi'\rangle_i=|4\rangle$), then the $N$ loops evolution operator is given by
$U_\mathrm{N}=\left ( \begin{array}{cc}
a_1& b_1\\
a_2& b_2
\end{array}\right),$
where $a_2=-b_1^*, b_2=a_1^*$ to keep $U_\mathrm{N}$ unitary. As shown in the SM~\cite{SM}, $|a_1|=2\langle\sigma_z\rangle-1$ and $\mathrm{Arg}(b_1)-\mathrm{Arg}(a_1)=\arctan(\langle\sigma_y\rangle/\langle\sigma_x\rangle)$, where $\mathrm{Arg}(x)$ gives the phase of the complex number $x$. Hence, $a_l, b_l (l=1,2)$ can be obtained by taking tomography of the pseudospin $\mathbf{F}$~\cite{DMethod}. In the SM~\cite{SM}, we show our tomograph is based no non-adiabatic holonomic quantum gates.

In Fig. 3(a), we drive the system evolves along the triangle loop with $\theta=0.5\pi$ and $\Delta\varphi=0.1\pi$.  It can be seen that $\mathbf{F}$ rotates around specific axis determined by $U^\mathrm{g}$ \cite{SM}. The experimental results of the first five loops are symbolled by the black stars in Fig. 3(a), where the data of the fifth loop are obtained as $\langle\sigma_x\rangle_{5}=0.334$, $\langle\sigma_y\rangle_{5}=0.299$ and $\langle\sigma_z\rangle_{5}=-0.894$. According to the formulation of $U_\mathrm{N}$, the evolution operator $U_5(0.5\pi,0.1\pi)$ of five loops can be obtained and the single loop evolution operator is obtained by $U(0.5\pi,0.1\pi)=U_5^{1/5}(0.5\pi,0.1\pi)$.

 We continue to extract the matrix of the gauge fields in the loop B-B'-C'-C-B  depicted in Fig. 1(c). We select the parameters $\Delta\theta=0.1\pi, \Delta\varphi=0.1\pi$. According to Eq.(5), the NAGFs are given by $\bar{F}_{\theta\varphi}(\theta_0, \varphi_0)=[U(\theta_0+\Delta\theta, \varphi_0+\Delta\varphi)-U(\theta_0, \varphi)]/\delta S'$ of which the evolution operators $U$ are traced by five loops evolution. The experimental results along the $\varphi_0=0$ longitude line are shown in Fig. 3(b). The experimental data appear to be sinusoidal and fitted well with the theoretical simulation. Here the non-vanishing non-diagonal elements $|\bar{F}_{\theta\varphi}^{12}|$ clearly show the non-Abelian characteristic of the measured gauge fields. As mentioned before, $\bar{F}_{\theta\varphi}$ is defined in the Cartesian coordinates with $\delta S'=\Delta\theta\Delta\varphi$. To recover the characteristic of SU(2) monopole (independent of the field angles), we refine the data by $\bar{F}^S_{\theta\varphi}=\bar{F}_{\theta\varphi}/\sin\theta$ which are plotted in Fig. 3(c).

{\sl Conclusions.--} In summary, we have theoretically proposed and experimentally demonstrated an approach to measure NAGFs. This method  offers the advantage of amplification through multi-loop evolution.
 Moreover, we can accelerate the manipulation by incorporating a shortcut to adiabaticity through the addition of an auxiliary Hamiltonian~\cite{Demirplak2003,Demirplak2005,Berry2009,Chen2010,Bason2012,Du2016}. Our results have the potential to be applied in various study of geometrical and topological phenomena in quantum systems.

\bigskip

\begin{acknowledgments}
 The authors thank Zhen-Yu Wang for helpful discussions. This work was supported by the Science and Technology Program of Guangzhou (Grant No. 202201010533), the China Postdoctoral Science Foundation(Grant No. 2022M721222), the Guangdong Basic and Applied Basic Research Foundation (Grants No. 2021A1515110668, No. 2023A1515011550), the National Key Research and Development Program of China (Grants No. 2022YFA1405300, No. 2022YFA1405303, and No. 2020YFA0309500), the National Natural Science Foundation of China (Grants No. 12074132, No. 12225405, No. 12247123, No. U20A2074, and No.12074180), and the Innovation Program for Quantum Science and Technology (Grant No. 2021ZD0301705).

Q.X.Lv, H.Z.Liu, and Y.X.Du contribute equally to this work.
\end{acknowledgments}

\appendix
\section{Supplemental Material:Measurement of non-Abelian gauge fields using multi-loop amplification}
In this Supplemental Material, we provide a more comprehensive discussion on the properties of the four-level Hamiltonian and offer additional experimental details.

\section{I. The degenerate eigen subspaces and the induced non-Abelian gauge fields}
The eigenstates of the Hamiltonian (3) in the main text correspond to the eigenvalues $\lambda_1=\lambda_2=-\Omega_0$ are given by
\begin{equation}
\begin{split}
&|D_1\rangle=\sin\frac{\theta}{2}|1\rangle-\cos\frac{\theta}{2}(\cos\phi e^{i\varphi}|2\rangle-\sin\phi|4\rangle),\\
&|D_2\rangle=-\sin\frac{\theta}{2}|3\rangle-\cos\frac{\theta}{2}(\sin\phi|2\rangle+\cos\phi e^{-i\varphi}|4\rangle),
\end{split}
\end{equation}
while the eigenstates correspond to the eigenvalues $\lambda_3=\lambda_4=\Omega_0$ are given by
\begin{eqnarray}
\begin{split}
&|B_1\rangle=-\sin\frac{\theta}{2}|2\rangle-\cos\frac{\theta}{2}(\cos\phi e^{-i\varphi}|1\rangle-\sin\phi|3\rangle),\\
&|B_2\rangle=\sin\frac{\theta}{2}|4\rangle-\cos\frac{\theta}{2}(\sin\phi|1\rangle+\cos\phi e^{i\varphi}|3\rangle).
\end{split}
\end{eqnarray}

  One significant advantage of this four-level configuration is that the degenerate subspace in such double-$\Lambda$ system with the lowest eigenvalue of $-\Omega_0$ will have a longer coherence time compared to the typical tripod system where the degenerate subspace is not the lowest energy level\cite{Zhang2008}. Under the subspace spanned by the lowest energy eigenstates $\{|D_1\rangle, |D_2\rangle\}$ one will obtain the non-Abelian gauge potentials as $\mathbf{A}=(A_\theta, A_\phi, A_\varphi)$ with
\begin{eqnarray}\nonumber
&&A_\theta=0,\\
&&A_\phi=-i\cos^2(\theta/2)\mathbf{\kappa}_\phi\cdot\mathbf{\sigma},\\ \nonumber
&&A_\varphi=i\cos^2(\theta/2)\cos\phi\mathbf{\kappa}_\varphi\cdot\mathbf{\sigma},
\end{eqnarray}
where $\mathbf{\sigma}=(\sigma_x, \sigma_y, \sigma_z)$ with $\sigma_{x,y,z}$ being the Pauli Matrixes. The unit vectors $\mathbf{\kappa}_\phi=(-\sin\varphi, \cos\varphi, 0)$ and $\mathbf{\kappa}_\varphi=(-\sin\phi\cos\varphi, \sin\phi\sin\varphi, \cos\phi)$. Components of the non-Abelian gauge fields along two of coordinates $\theta, \varphi, \phi$ are given by
\begin{eqnarray}\nonumber
&&F_{\theta\varphi}=-i\sin\theta\cos\phi\mathbf{\kappa}_{\theta\varphi}\cdot\mathbf{\sigma}/2,\\
&&F_{\theta\phi}=\sin\theta\mathbf{\kappa}_{\theta\phi}\cdot\mathbf{\sigma}/2,\\ \nonumber
&&F_{\phi\varphi}=2\cos^2(\theta/2)(1-\cos^2(\theta/2))\cos\phi\kappa_{\phi\varphi}\cdot\sigma,
\end{eqnarray}
where the unit vectors   $\mathbf{\kappa}_{\theta\varphi}=(\sin\phi\cos\varphi, \sin\phi\sin\varphi, \cos\phi)$, $\mathbf{\kappa}_{\theta\phi}=(\sin\varphi, -\cos\varphi, 0)$, and $\mathbf{\kappa}_{\phi\varphi}=(\cos\phi\cos\varphi, \cos\phi\sin\varphi, \sin\phi)$. Similar results can be found in the upper eigen subspace $\{|B_1\rangle, |B_2\rangle\}$. We will focus on the lower eigen subspace in the following discussion.

\renewcommand{\thefigure}{S1}
\begin{figure*}[ptb]
\begin{center}
\includegraphics[width=16cm]{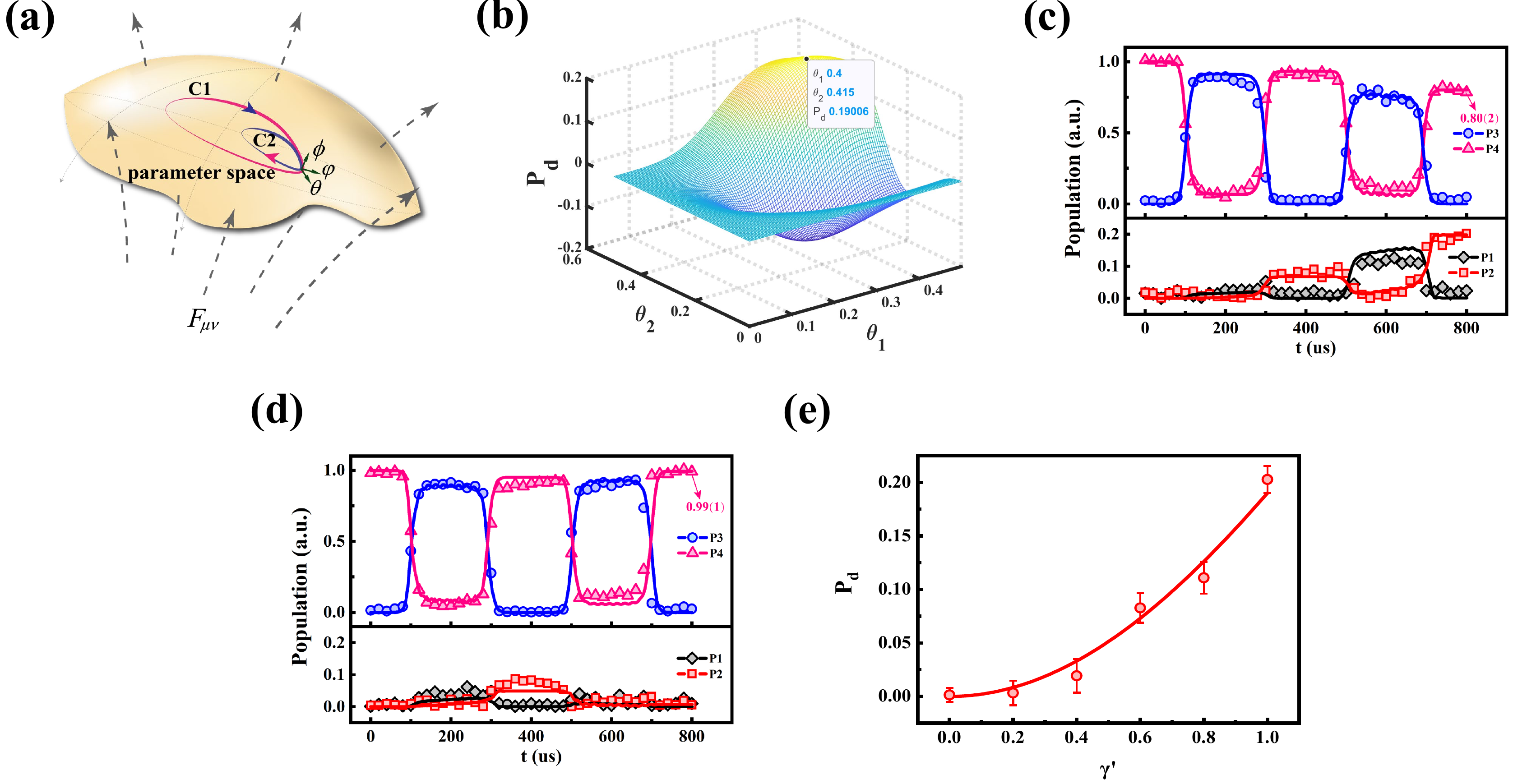}
\caption{Measurements of non-commutativity of non-Abelian gauge fields. (a) A quantum system undergoing cyclic and adiabatic evolution in the parameter space will acquire a Wilzeck-Zee phase that is associated with a non-Abelian gauge field $F_{\mu\nu}$. The commutativity of the gauge field can be detected by observing the final population difference resulting from changing the order of two successive closed loops.
 (b) Population difference versus parameters $\theta_1$ and $\theta_2$ in $\mathrm{C}_1$ and $\mathrm{C}_2$, respectively. (c)(d) population dynamics for evolution $\textrm{C}_1\textrm{C}_2$ and $\textrm{C}_2\textrm{C}_1$ at $\gamma'=1$, respectively. Pink triangles, blue circles, red squares, gray diamonds: experimental results for population at each level $|i\rangle (P_i), i=1,2,3,4$ which have been averaged for 5 measurements. Solid lines: simulation results. (e) Difference of final population left in $|4\rangle$ between composite loops $\textrm{C}_1\textrm{C}_2 (\textrm{P}_{12})$ and $\textrm{C}_2\textrm{C}_1 (\textrm{P}_{21})$. Non-vanishing population difference corresponds to non-Abelian characteristic while vanishing population difference corresponds to the Abelian one. The gauge fields can change from non-Abelian to Abelian by tilting the parameter $\gamma'$. Red Circles with error bars: experimental data which have been averaged by 10 measurements; Red solid line: simulation curve. }
\end{center}
\end{figure*}

The Hamiltonian (3) in the main text can be experimentally realized in $^{87}$Rb cold atomic system, where we encode four Zeeman sublevels in the ground states by $|1\rangle=|F=2,m_F=-1\rangle, |2\rangle=|F=1,m_F=-1\rangle, |3\rangle=|F=2,m_F=0\rangle, |4\rangle=|F=1,m_F=0\rangle$, respectively. A magnetic field about 0.5 Gauss is applied to resolve the degenerate of the Zeeman levels. $|1\rangle, |2\rangle$ and $|3\rangle, |4\rangle$ are coupled by the $\pi-$ transition microwaves while $|1\rangle, |4\rangle$ ($|2\rangle, |3\rangle$) are coupled by the $\sigma^--$ ($\sigma^+-$) transition microwaves, respectively. The Rabi frequencies, frequencies, and phases of microwaves are all adjustable which offer a fully controllable Hamiltonian. The system is initialized to state $|4\rangle$ which is realized by optically pumping atoms to $|3\rangle$ and then being transferred to $|4\rangle$ by microwaves with $\pi$ pulse. A more detailed description of the experimental setup can be seen in Ref.\cite{Lv2021}.

The non-Abelian gauge fields can be induced in the degenerate eigen subspaces. To confirm the generation of non-Abelian gauge field in this four-level system, it is necessary to examine its major characteristic, which is non-commutativity. The non-commutativity of $F_{\mu
\nu}$ can be detected by analyzing the corresponding cyclic evolution operator $U^{\mathrm{c}}$ associated with $A_{\mu}$~\cite{Zhang2008},

\begin{equation}
U^{\mathrm{c}}=\left ( \begin{array}{cc}
U^{11} & U^{12}\\
U^{21} & U^{22}
\end{array}\right)=\hat{P}e^{-\oint_\mathrm{c}A_\mu d\mu},
\end{equation}
where $\hat{P}$ represents the path-ordered operator. As shown in Fig. S1(a), two different loops $\textrm{C}_1$ and $\textrm{C}_2$ formed by vector $\mathbf{R}(\theta,\phi,\varphi)$ in parameters space are adopted. Assuming that the evolution operators of $\textrm{C}_1$ and $\textrm{C}_2$ are given by $U_1$ and $U_2$, respectively, then the  evolution operators of composite paths with order $\textrm{C}_1\textrm{C}_2$ and counter-order $\textrm{C}_2\textrm{C}_1$ will be given by $U_{\mathrm{o}}=U_2U_1$ and $U_{\mathrm{co}}=U_1U_2$, respectively. $U_2U_1\neq U_1U_2$ can be validated by the non-diagonal matrix elements $|U_{\mathrm{o}}^{12(21)}|^2\neq|U_{\mathrm{co}}^{12(21)}|^2$, namely, the population transfers after evolution. It should be noted that the starting point of $C_1$ and $C_2$ must coincide, otherwise, two loops with disconnected starting point may be treated as two non-commutative gates in geometric quantum computation.

To realize two different loops in parameter space, the parameters $\theta, \phi, \varphi$ of $\mathrm{C}_1, \mathrm{C}_2$ can be specified as $\theta^{(1(2))}=\theta_{1(2)}f(t)$, $\phi^{(1)}=\phi^{(2)}=\gamma f(t)$, $\varphi^{(1)}=2\pi t/T$, and $\varphi^{(2)}=2\pi f(t)t/T$, where $T$ is the evolution period. By setting $f(t:0\rightarrow T):0\rightarrow 0$, the Hamiltonian (3) in the main text satisfies $H(0)=H(T)$ which experiences cyclic evolution in parameter space. Here we choose to tilt $\theta, \phi, \varphi$ simultaneously since $F_{\theta\varphi}, F_{\theta\phi}, F_{\phi\varphi}$ are all non-vanishing.

One of the primary challenges in testing non-commutativity is the need to detect the population of eigenstates, as the gauge fields are defined within their respective subspaces. However, this obstacle can be overcome by selecting appropriate parameter settings.
It is evident from Eqs. (S1) and (S2) that, with the provided loops $\mathrm{C}_1$ and $\mathrm{C}_2$, the association between eigenstates and bare states can be elucidated as follows:

\begin{eqnarray}
\begin{split}
&|D_1(t=0)\rangle=|D_1(t=T)\rangle=|2\rangle\\
&|D_2(t=0)\rangle=|D_2(t=T)\rangle=|4\rangle\\
&|B_1(t=0)\rangle=|B_1(t=T)\rangle=|1\rangle\\
&|B_2(t=0)\rangle=|B_2(t=T)\rangle=|3\rangle.
\end{split}
\end{eqnarray}
Hence, by detecting the population of the bare states under suitable conditions, one can easily detect the population of the eigenstates, thereby significantly reducing the experimental challenge.

The experiments conducted to examine non-commutativity are carried out as follows: Firstly, we prepare the initial state of the system to be $|\Psi\rangle_i=|4\rangle=|D_2\rangle$. Next, we drive the system to evolve along the closed paths $\mathrm{C}_1$ and $\mathrm{C}_2$, and measure the population $P_{12}=|U_o^{12}|^2$ transferred to $|2\rangle=|D_1\rangle$.
On the other hand, we drive the system to evolve along the closed paths $\mathrm{C}_2$ and $\mathrm{C}_1$, and measure the population $P_{21}=|U_{co}^{12}|^2$ transferred to the state $|2\rangle$. For more detailed information on how to measure the population in the four-level system, please refer to Ref.\cite{Lv2021}.  If the population difference $P_d=P_{12}-P_{21}\neq0$, then $U_o\neq U_{co}$.

Just as the theoretical simulation shown in Fig. S1(b), we have set $\theta_1=0.4\pi$, $\theta_2=0.415\pi$, and $\gamma=\pi/16$ to achieve maximum population difference and non-commutativity.  Please note that all evolution must be adiabatic to ensure that the system remains in the lower subspace where the Rabi frequency $\Omega_0=2\pi\times50$kHz and $T=400\mu$s, satisfying the adiabatic condition $\Omega_0 T=40\pi\gg2\pi$. The population dynamics during the evolution $\mathrm{C}_1\mathrm{C}_2$ and $\mathrm{C}_2\mathrm{C}_1$ are illustrated in Fig. S1(c)(d), respectively.
 The pink triangles, blue circles, red squares, and gray diamonds represent the experimental data for population $P_i$ at level $|i\rangle$, where $i=1,2,3,4$. Meanwhile, the solid lines depict the simulation results. The final population of $|2\rangle$ is $0.80(2)$ for $\mathrm{C}_1\mathrm{C}_2$ and $0.99(1)$ for $\mathrm{C}_2\mathrm{C}_1$, respectively.

 Hence, it has been established through non-commutativity that the induced gauge fields must be non-Abelian. Additionally, the properties of the induced gauge fields can be modified by varying $\gamma'=\gamma/(\pi/16)$. In Figure S1(e), it can be observed that when $\gamma'=0$, the four-level system is reduced to two two-level systems, constructed by $\{|1\rangle, |2\rangle\}$ and $\{|3\rangle, |4\rangle\}$, respectively. Each of two subsystems induces an Abelian field and thus there is no population difference. On the other hand, when $\gamma'=1$, we arrive at the case depicted in Fig. S1(c)(d). The total variation of $P_d$ during the interval $\gamma'\in[0,1]$ is illustrated in Figure S1(e).

\section{II. Relationship between gauge fields and the geometric gate in the orange-slice model}
In this section we show the derivation of the geometric gate (6) in the main text. We assume the evolution operator along closed path in parameter space with parameters $(\theta, \varphi)$ to be
\begin{equation}
U^\mathrm{g}=\left ( \begin{array}{cc}
\cos\kappa e^{-i\alpha} & i\sin\kappa e^{-i\beta}\\
i\sin\kappa e^{i\beta} & \cos\kappa e^{i\alpha}
\end{array}\right).
\end{equation}
Substituting (S7) into the Sch\"{o}rdinger's equation and expanding the quantum state by $|\Psi\rangle=c_1|D_1\rangle+c_2|D_2\rangle$, we have
\begin{equation}
\dot{U}^\mathrm{g}=-A_\varphi\dot{\varphi}U^\mathrm{g},
\end{equation}
where the dot means derivation over time $t$. By adopting the triangle loop model in Fig. 1(c) in the main text ($\dot{\varphi}=0$ except for $t=T/2$), we will obtain
\begin{eqnarray}
\begin{split}
&\beta=-\alpha=\cos^2\phi\oint\cos^2(\theta/2)d\varphi=\cos^2\phi\Delta\varphi,\\
&\kappa=\cos\phi\sin\phi\oint\cos^2(\theta/2)d\varphi=\cos\phi\sin\phi\Delta\varphi,
\end{split}
\end{eqnarray}
where $\Delta\varphi$ is the  angle of the orange-slice model. One can further deduce
\begin{equation}
U^\mathrm{g}=e^{i\eta \vec{h}\cdot\sigma},
\end{equation}
where $\vec{h}=(\sin\chi\cos\xi,\sin\chi\sin\xi,\cos\chi)$, $\xi=\beta$, $\tan\chi=\tan\kappa/\sin\beta$, and $\sin\eta=\sqrt{\sin^2\kappa+\cos^2\kappa\sin^2\beta}$. The evolution can be viewed as the quantum state circling around the axis $\vec{h}$ by an angle of $\eta$ once.
 (and $N\eta$ for $N$ triangle loops).

\setcounter{figure}{0}
\renewcommand{\thefigure}{S2}
\begin{figure}[ptb]
\begin{center}
\includegraphics[width=6.5cm]{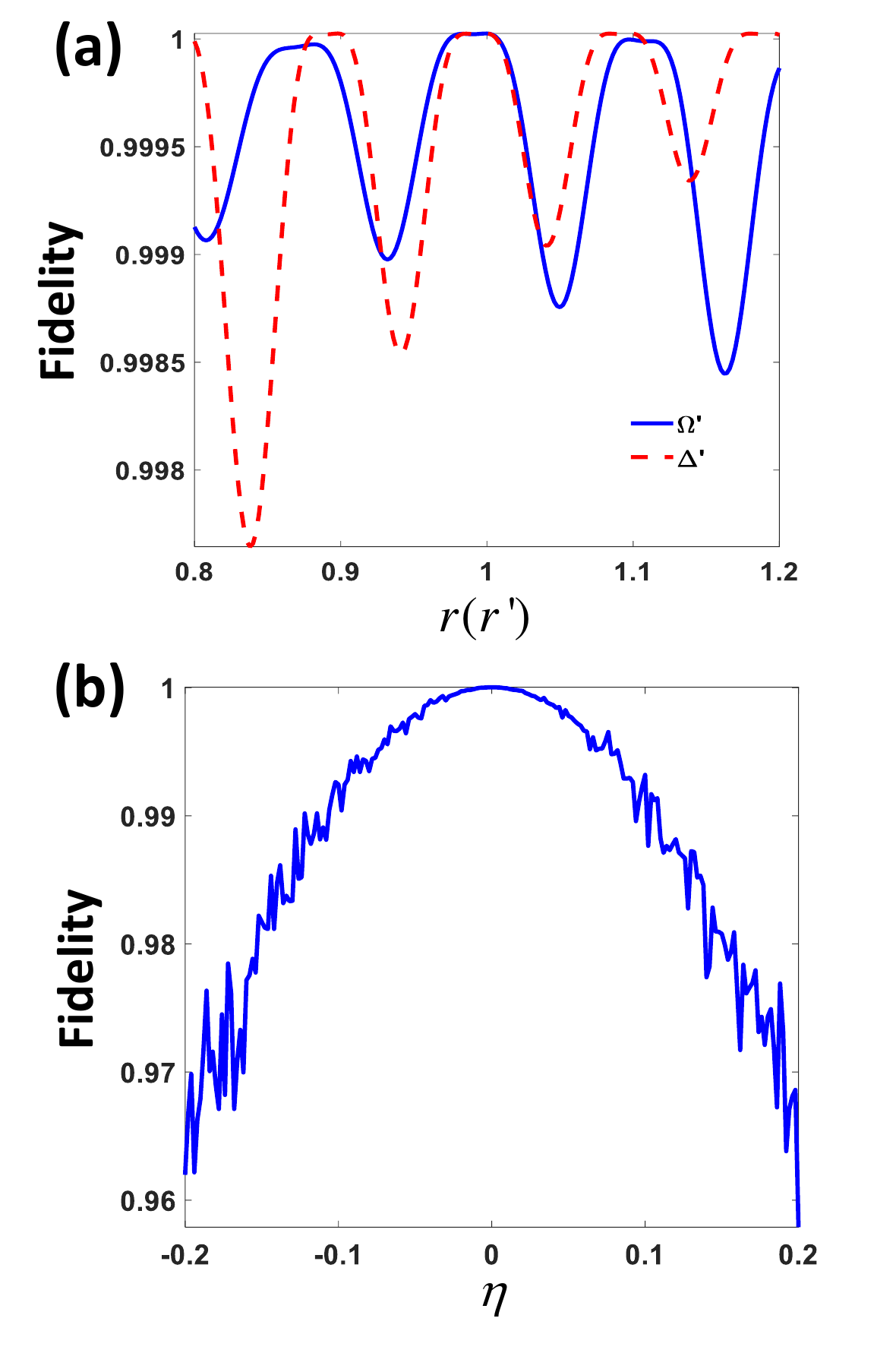}
\caption{Robustness of the evolution operator against two types of errors: (a) systematic errors and (b) random noise.
}
\end{center}
\end{figure}

When $\Delta\varphi\ll\pi$, we have $\sin\kappa\sim\kappa$ and $\cos\kappa\sim0$, thus Eq.(S7) modifies to
\begin{equation}
U^\mathrm{g}=\left ( \begin{array}{cc}
1-i\beta & i\kappa e^{-i\beta}\\
i\kappa e^{i\beta} & 1+i\beta
\end{array}\right).
\end{equation}
Meanwhile, according to Eq.(2) in the main text, we have $U'=1-\bar{F}_{\theta\varphi}(\Delta\varphi)\Delta S=1-\int\int F_{\theta\varphi}d\theta d\varphi$ which can easily recover the result of Eq.(S9) for $\Delta\varphi\ll\pi$ and $\phi\ll\pi$.
Therefore, by measuring the evolution of the orange-slice model, specifically the geometric gate, we can detect the non-Abelian gauge field. It is important to note that although Eq. (S7) is solved by expanding the quantum state with eigenstate basis, we only need to detect the bare states to perform tomography under the condition that $\theta(0)=\theta(T)=0$.
In the case of the gauge field in the square loops discussed in the main text, we are able to measure the evolution of two distinct triangle loops that encompass the square loop. We can extract the gauge field strength with the difference between these two loops. This is possible due to the linearity of Eq. (2) in the main text.


\setcounter{figure}{0}
\renewcommand{\thefigure}{S2}
\begin{figure}[ptb]
\begin{center}
\includegraphics[width=8cm]{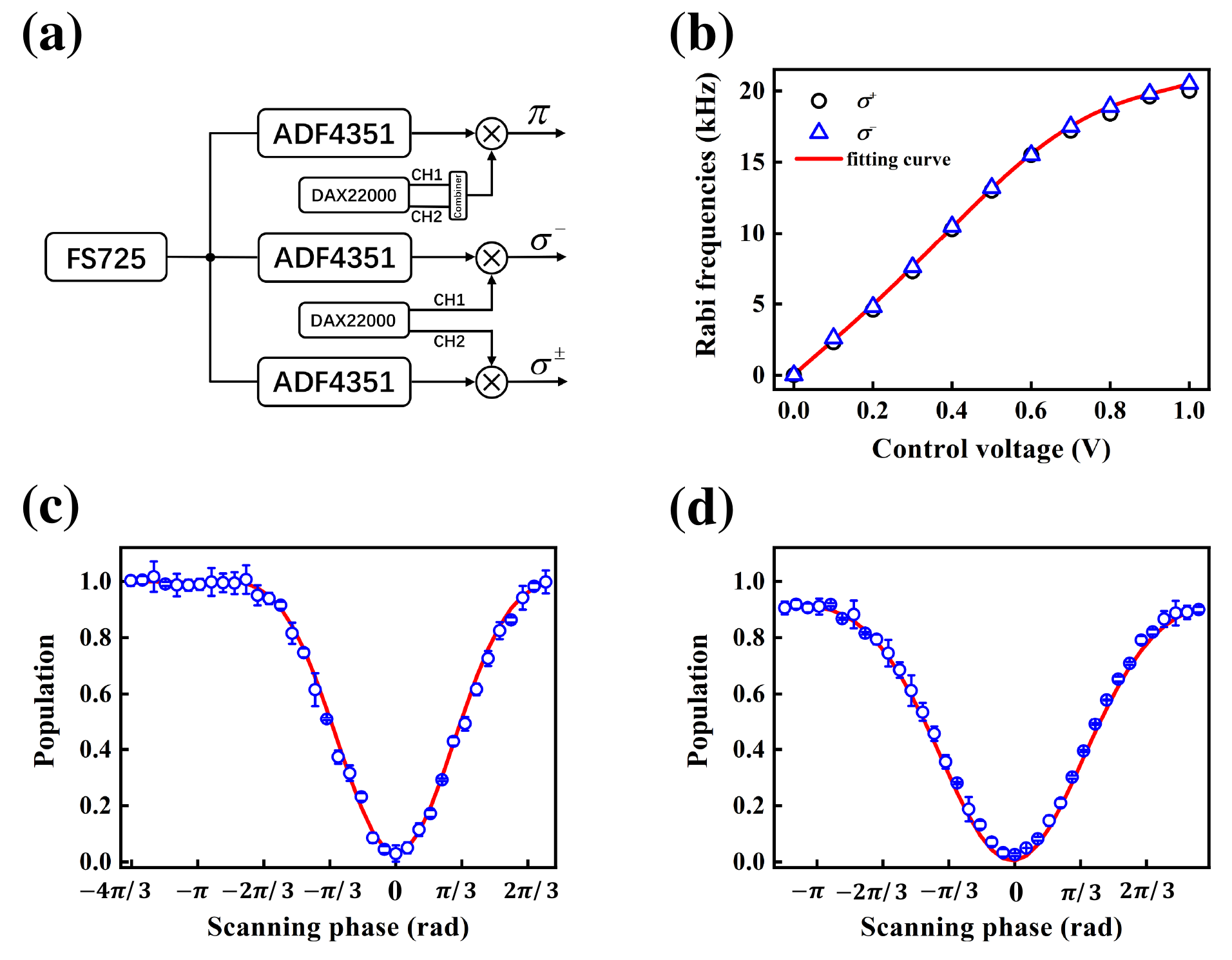}
\caption{Calibrating the phases and amplitudes of the microwaves. (a) Electronic circuits of the microwave system. To keep phase coherence between different microwaves, all eigen sources are synchronized to the same atomic clock. FS725: atomic clock. ADF4351: eigen microwave sources. DAX22000: two channels arbitrary waveform generator (CH1 and CH2). By mixing the signals from eigen sources and radio sources, arbitrary time-varying amplitude, frequency, and phase of the microwave field can be achieved. (b) Rabi frequencies of the intermediate coupling $g$ controlled with $\sigma^\pm$ . Black circles: $\sigma^+$. Blue triangles: $\sigma^-$. Red solid line: fitting curve with a polynomial. The Rabi frequencies of $\sigma^+$ and $\sigma^-$ can be adjusted to be equivalent at varying control voltages. (c)(d) Population difference by scanning the relative phases between two  microwave pulses. (c) $\sigma^\pm$ versus $\pi$. (d) $\sigma^-$ versus $\pi$. Red solid lines: fitting curves with the form $f(\phi_{ab})=\sqrt{\Omega_a^2+\Omega_b^2+2\Omega_a\Omega_b\cos{\phi_{ab}}}$, where $\Omega_a$ and $\Omega_b$ are the Rabi frequencies and $\phi_{ab}$ is the relative phase. Blue dots with error bars: experimental data.
}
\end{center}
\end{figure}

In order to demonstrate the resilience of our measurement of the evolution operator against systematic errors and random noise,  we evaluate the fidelity and the results are plotted in Fig. S2(a) and S2(b). The fidelity is defined as the absolute value of the inner product between the actual state, $|\Psi_{act}\rangle$, and the ideal state, $|\Psi_{ideal}\rangle$, denoted as $F=|\langle \Psi_{act}|\Psi_{ideal}\rangle|$. We present the fidelity plotted against the deviation in Rabi frequency (blue-solid line) and detuning (red-dashed line) in Fig. S2(a), which are induced by $\Omega'=r\Omega$ and $\Delta'=r'\Delta$, respectively. It is evident that the fidelity is greater than 0.99 for both cases and is resilient to systematic errors. The fidelity versus random noise in Rabi frequency is depicted in Fig. S2(b), where the noise is introduced by $\Omega'=(1+ \eta\times rand)\Omega$, with $rand$ representing a random noise function with a mean value of zero. The theoretical results in Fig. S2(b) have been averaged over 30 trials and demonstrate that the fidelity remains robust against random noise across a wide range.

\section{III. Experimental setup}

The experimental setup is shown in Fig. 1(b) in the main text. Cold atoms are trapped in an optical dipole trap and cooled down to a temperature of $10\mu K$ through evaporation. Three microwave horn antennas (Here $\sigma^--$polarized microwave horn antenna is omitted, which is used for initial state preparation and quantum state tomography) are employed to couple different levels of ${}^{87}\mathrm{Rb}$. The two microwaves, which are individually $\pi$-polarized, originate from a single linearly polarized microwave horn antenna. These microwaves are utilized to realize $\pi$-transitions (that is, the transition between $|1\rangle$($|3\rangle$) and $|2\rangle$($|4\rangle$)).
 Two additional $\sigma^\pm$-polarized microwaves, which stem from the $\sigma^+$-polarized microwave horn antenna, are not considered independent. This is because the antenna is not completely pure and there is back reflection from the surrounding environment, resulting in the production of $\sigma^-$-polarized microwaves from the same antenna. Therefore, the $\sigma^{\pm}$ transition of the intermediate coupling, which refers to the transition between $|2\rangle$($|4\rangle$) and $|3\rangle$($|1\rangle$), can be achieved using only one circular polarization horn antenna. As shown in Fig. S2(a), the microwave eigen sources are generated by frequency synthesizers (Analog Devices, ADF4351) that are all connected to the same atomic clock. All horn antennas, including those polarized as $\pi$, $\sigma^\pm$, and $\sigma^-$, emit microwaves that are generated from a device (Waveponds, DAX22000-8M) through a process of mixing radio frequencies  with their eigen frequencies.

We introduce intermediate couplings between the two single-qubits($|1\rangle,|4\rangle$ and $|2\rangle,|3\rangle$) to induce non-Abelian gauge fields. $\sigma^+$ and $\sigma^--$transition in the intermediate coupling should be equal according to the  Hamiltonian (3), which can be realized by regulating the position of the $\sigma^\pm$ horn antenna.
 We prepared the system to be in the state $\left|{2}\right\rangle$ or $\left|{4}\right\rangle$ and measured the Rabi oscillation caused by the $\sigma^+$ and $\sigma^-$ transition microwaves, respectively. Based on the measured Rabi frequencies, we meticulously adjusted the position of the $\sigma^\pm$ horn antenna until the $\sigma^+$ and $\sigma^-$ transitions in the intermediate couplings were equal. The data for $\sigma^+$ and $\sigma^-$ transitions are presented in Figure S3(b), which clearly illustrates that the Rabi frequencies of $\sigma^+$ (represented by black circles) are equivalent to those of $\sigma^-$ (represented by blue triangles) across varying control voltages. This confirms the successful achievement of intermediate coupling.

\section{IV.Calibration of the relative phases}

To achieve the four-level Hamiltonian (3), precise calibration of the relative phases among the microwaves is essential. During the experiment, phase differences may arise due to various factors such as transport paths, electronic delays, and the radio frequencies used for initializing the microwave phases. To address this issue, we detect the relative phases of the microwaves by observing the interference between the Rabi frequencies of the corresponding microwaves. We begin by preparing the system in the initial state $|4\rangle$ and tuning two of the microwaves to be resonant with both $|4\rangle$ and $|3\rangle$. The resulting induced population is dependent on the synthetic Rabi frequency, which is determined by the phase difference between the Rabi frequencies of the two microwaves. By measuring the population, we can extract the relative phases between the microwaves.

In Fig.S3(c) and (d), the relative phases between the $\pi$-transition microwave and one of the $\sigma^\pm(\sigma^-)$-transitions are presented. The lowest population corresponds to zero phase difference. The experimental data is represented by blue circles with error bars, while the theoretical curves are depicted by red solid curves in the form of $f(\phi_{ab})=\sqrt{\Omega_a^2+\Omega_b^2+2{\Omega_a}{\Omega_b}\cos(\phi_{ab})}$. The Rabi frequencies are denoted by $\Omega_a$ and $\Omega_b$, and the relative phase is indicated by $\phi_{ab}$. Finally, we adjust the phase of one microwave, specifically the $\pi-$transition microwave, and then align the phases of the $\sigma^\pm$ and $\sigma^-$-transition microwaves to meet the required phase conditions. We note that the two channels of the $\pi$-transition microwave are produced from the same source and transported along identical paths, resulting in their phases being automatically synchronized.

\section{V. Quantum state tomography and the measurement errors}

\setcounter{figure}{0}
\renewcommand{\thefigure}{S4}
\begin{figure}[ptb]
\begin{center}
\includegraphics[width=4cm]{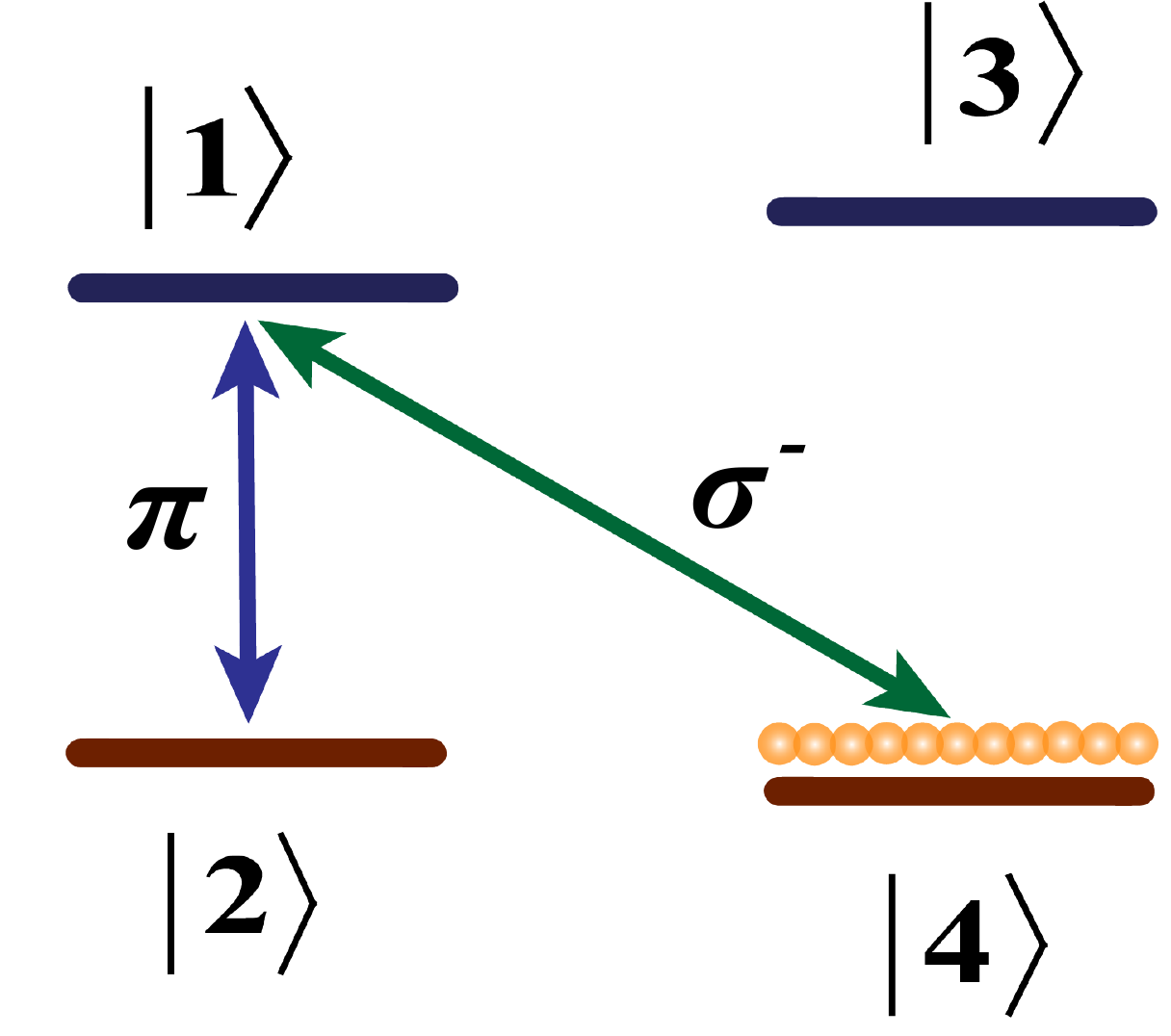}
\caption{Quantum state tomography based on non-adiabatic holonomic single-qubit gates. Bare states ${\left|{1}\right\rangle}, {\left|{2}\right\rangle}$ and ${\left|{4}\right\rangle}$ form a $\Lambda$ configuration where $\left|{1}\right\rangle$, $\left|{2}\right\rangle$ and $\left|{1}\right\rangle$, $\left|{4}\right\rangle$ are coupled by the $\pi-$ and $\sigma^--$ transition microwave, respectively.
}
\end{center}
\end{figure}

\setcounter{figure}{0}
\renewcommand{\thefigure}{S5}
\begin{figure}[ptb]
\begin{center}
\includegraphics[width=5cm]{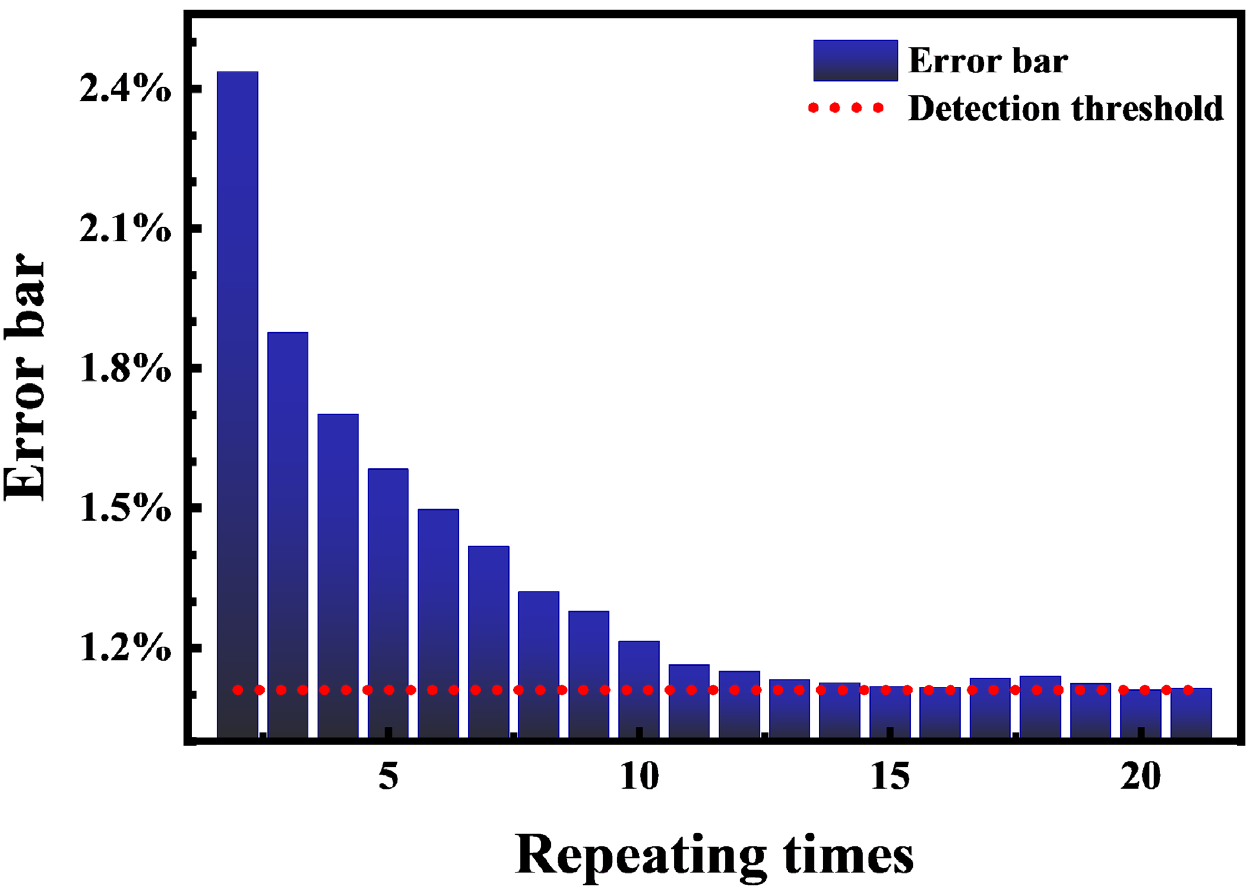}
\caption{The error bars on the experimental data decrease as the number of measurements increases. As the number of measurements approaches a certain threshold, the error bars converge towards a value of 0.011. This value can be used as a criterion for determining the minimum detectable signal.
}
\end{center}
\end{figure}

In this section, we will introduce how to obtain tomography of the pseudospin spanned by $|2\rangle$ and $|4\rangle$. By utilizing the complete control of the system through microwaves, we will utilize the non-adiabatic holonomic single-qubit gates to achieve quantum state tomography. We consider a three-level system composed of $\{|2\rangle, |1\rangle, |4\rangle\}$ as illustrated in Fig.S4. The interaction Hamiltonian in the rotating frame will take the following form~\cite{NJP2012}
\begin{align}
\label{H_NHQC}
H_{\mathrm{nd}}= {\Omega _{21}}(t){e^{i{\varphi _2}}}|2\rangle \langle 1| + {\Omega _{41}}(t){e^{i{\varphi _4}}}|4\rangle \langle 1| + {\rm{H}}.{\rm{c}}.
\end{align}
By parameterizing $\Omega _{21}=\Omega _{\mathrm{nd}}\sin\theta_{\mathrm{nd}}, \Omega _{41}=\Omega _{\mathrm{nd}}\sin\theta_{\mathrm{nd}}$ with $\varphi_{nd}=\varphi_2-\varphi_4+\pi$, $\Omega_{\mathrm{nd}}=\sqrt{{\Omega_{21}}{{(t)}^2}+{\Omega_{41}}{{(t)}^2}}$, the dynamics can be considered as the resonant coupling between the states $|b\rangle=\sin(\theta_{\mathrm{nd}}/2)e^{-i\varphi}|2\rangle-\cos(\theta_{\mathrm{nd}}/2)|4\rangle$ and $|1\rangle$, while the dark eigenstate $|d\rangle=\cos(\theta_{\mathrm{nd}}/2)|2\rangle+\sin(\theta_{\mathrm{nd}}/2)e^{i\varphi}|4\rangle$ is decoupled. Hence, when the condition for cyclic evolution is satisfied (i.e., $\int_0^\tau{\Omega_{\mathrm{nd}}dt}=\pi$, where $\tau$ is the evolution period), a specific single-qubit gate can be obtained for the basis states $|2\rangle$ and $|4\rangle$, as shown by:

\begin{equation}
\begin{split}
\label{U_NHQC}
U_{\mathrm{tomo}} &= \left( {\begin{array}{*{20}{c}}
{\cos \theta_{\mathrm{nd}} }&{\sin \theta_{\mathrm{nd}} {e^{ - i\varphi }}}\\
{\sin \theta_{\mathrm{nd}} {e^{i\varphi }}}&{ - \cos \theta_{\mathrm{nd}} }
\end{array}} \right).\\
\end{split}
\end{equation}
By choosing specific parameters in Eq.(\ref{U_NHQC}), i.e., $\theta_{nd}=\pi/2, \varphi_{nd}=0$, $\sigma_y$-operation can be realized, while $\theta_{nd}=\pi/2,\varphi_{nd}=\pi/2$ correspond to $\sigma_x$-operation.
In our experiment, we need to measure three components of the Bloch vector, namely $\langle\sigma_{{ x,y,z} }\rangle$. The expectation value of $\langle\sigma_z\rangle$ can be determined by measuring the atomic population difference between states $|2\rangle$ and $|4\rangle$.
In order to measure the expectation value of Bloch vectors $\langle\sigma_x\rangle$ and $\langle\sigma_y\rangle$, a $\pi/2$ microwave pulse is applied with a phase of either 0 or $\pi/2$. This pulse is used to rotate the Bloch vector accordingly before the population difference measurement is taken. We now discuss measurement errors. As depicted in Fig.S5, the error bars for population $P_4$ decrease as the number of measurements increase. Eventually, the error bars converge to a value of $0.011$, which can be considered as the threshold for detecting the smallest possible signal.


\begin{thebibliography}{99}
\bibitem{Final_Population} The final population in states $|1\rangle$ and $|3\rangle$ is vanished according to the manipulation.
\end{thebibliography}
\end{document}